\documentstyle[12pt,psfig]{article}
\def\reference{\parskip 0pt\par\noindent\hangindent 0.5 truecm}
\def\kms{km ${\rm s}^{-1}$}

\def\lapp{\stackrel{<}{_{\sim}}}
\begin{document}
%
%
\title{Finding Pulsars at Parkes}
%


\author{R. N. Manchester
} 

\date{}
\maketitle

{\center
Australia Telescope National Facility, CSIRO, P.O. Box 76, Epping NSW
1710\\rmanches@atnf.csiro.au\\[3mm]
}

%
\begin{abstract}
There are many reasons why it is important to increase the number of known
pulsars. Not only do pulsar searches continue to improve statistical
estimates of, for example, pulsar birthrates, lifetimes and the Galactic
distribution, but they continue to turn up interesting and, in some cases, unique
individual pulsars. In the early days of pulsar astronomy, the Molonglo
radio telescope led the world as a pulsar detection instrument. However, the
Parkes radio telescope, with its frequency versatility and greater tracking
ablility, combined with sensitive receivers and powerful computer detection
algorithms, is now the world's most successful telescope at finding
pulsars. The Parkes multibeam survey, begun in 1997, by itself will come
close to doubling the number of known pulsars. Parkes has also been very
successful at finding millisecond pulsars, especially in globular
clusters. One third of the known millisecond pulsars have been found in just
one cluster, 47 Tucanae.
\end{abstract}


\bigskip

%
%

\section{Introduction}
Since the discovery of the first pulsar by Jocelyn Bell and Tony Hewish in
1967 (Hewish et al. 1968), many observatories throughout the world have
undertaken searches for these fascinating objects. Prior to the commencement
of the Parkes multibeam survey in mid-1997, these searches had resulted in
the discovery of about 750 pulsars. All but a few of these lie within our
Galaxy -- the only known extra-galactic pulsars are associated with the
Magellanic Clouds -- and all but a few have been discovered at radio
wavelengths. Most are relatively close on a Galactic scale -- their median
distance from the Sun is only about 3.5 kpc. This is not because pulsars are
clustered about the Sun. It is purely a result of the rather low radio
luminosity of pulsars, which makes them difficult to detect at large
distances.

It is now almost universally accepted that pulsars are neutron stars, with the
basic periodicity defined by rotation of the star.  The pulse period is
very predictable, but it is not constant. Despite various short-term
fluctuations observed to a greater or lesser extent in most pulsars, in the
long term the intrinsic period of all pulsars increases with time. Pulsars
are powered by the kinetic energy of rotation. They steadily lose
energy, mainly in the form of a high-energy wind of charged particles and
magnetic-dipole radiation, that is, electromagnetic waves at the
neutron-star rotation frequency.  

Pulsars may be divided into two main groups, based on the pulse period and the
rate at which it increases.  The first group, often called `normal'
pulsars, typically have periods of between 0.05 and 5 sec, and
characteristic ages, defined by $\tau_c = P/(2 \dot P)$, where $P$ is the
pulsar period and $\dot P$ is its secular rate of change, in the range
$10^3$ to $10^7$ years. The other group are the `millisecond' pulsars
(MSPs), most of which have periods of between 1.5 and 25 ms. A key property
of MSPs is their great age, typically between $10^8$ and $10^{10}$
years. Another key property is that most MSPs are members of a binary
system, in an orbit with another star. These properties suggest that MSPs
are in fact `recycled' neutron stars, spun up by accretion from a binary
companion. For an extensive review of MSP formation mechanisms and their
relation to X-ray binary systems, see Bhattacharya \& van den Heuvel (1991).

Not long after the discovery of the first MSP by Backer et al. (1982) it was
realised that the cores of globular clusters were a favourable environment
for the formation of MSPs (Hamilton, Helfand \& Becker 1985) . Several
groups began searches toward globular clusters and this effort was rewarded
by the discovery of the 3-ms pulsar PSR B1821$-$24 in the core of M28 by
Lyne et al. (1987). Over the next few years, more than 30 MSPs were
discovered in globular clusters. Of these, M15, with 8 pulsars (Wolszczan et
al. 1989, Anderson 1992), and 47 Tucanae with 11 (Manchester et al. 1991,
Robinson et al. 1995) stand out as the most prolific.

With 750 pulsars already known, why bother to find more? 

There are many good reasons.  Even though 750 sounds like a large number,
when you divide them into luminosity, distance and/or age bins, the number
in some bins is not all that large. In particular, low-luminosity pulsars
dominate the Galactic birthrate (e.g., Lyne et al. 1998) and yet we have a
rather small sample of them, leading to large statistical uncertainties
in birthrate calculations. Similarly, we know of very few pulsars at
distances comparable to the Galactic Centre, so estimates of the Galactic
population of pulsars are very uncertain except in the Solar neighbourhood.

An increased sample of pulsars also is of great value to timing
investigations and studies of the emission process. Young pulsars are known
to suffer glitches, that is, sudden increases in spin rate, and various
other forms of period irregularities. These phenomena are believed to result
from transitions in the superfluid interior of the neutron star and are one
of the few ways that we have of investigating the physics of ultra-dense
matter (Alpar, Cheng \& Pines 1989). However, the number of known
glitching pulsars is relatively modest (Wang et al. 2000) and there is
great variety in glitch properties. Similarly, there is a wide variety of
pulse emission properties, with special groups of pulsars such as those with
interpulses or wide profiles, high polarisation, drifting sub-pulses or null
periods. An increased sample is very useful for studies of properties such
as these.

Pulsars are also excellent probes of the interstellar medium (ISM). They are
pulsed, allowing measurement of the dispersive delay due to free electrons
in the ISM, and hence the column density of electrons along the path,
commonly expressed as a dispersion measure (DM) in the units cm$^{-3}$
pc. Given a model for the interstellar free-electron distribution
(e.g. Taylor \& Cordes 1993) pulsar distances can be estimated from their
DM.  Compared to most celestial radio sources, pulsars have strong linear
polarisation, and hence measurement of Faraday rotation is relatively
easy. Pulsars have the unique advantage that the DM along the path is also
known, so the mean line-of-sight magnetic field strength can be directly
estimated, leading to models for the Galactic magnetic field (Han,
Manchester and Qiao 1999). Pulsars also have the almost unique property that
they are of very small angular size, making possible observation of the full
range of effects due to scattering by small-scale fluctuations in the
interstellar electron density (Rickett 1990). Pulsars are also excellent
probes of the interstellar neutral gas (Frail et al. 1994). Most of these
investigations are limited by the spatial density of known pulsars in the
Galaxy, and so increasing the sample is of great value.

One of the most fascinating things about pulsars is the fact that pulsar
surveys keep turning up new and sometimes totally unexpected classes of
object. Even the discovery of the first pulsar itself was
serendipitious. Outstanding examples are the discovery of the Vela and Crab
pulsars (Large, Vaughan \& Mills 1968, Staelin \& Reifenstein 1968), the
first binary pulsar, PSR B1913+16 (Hulse \& Taylor 1974), the first
millisecond pulsar (Backer et al. 1982), the first globular cluster
pulsar (Lyne et al. 1987), the first eclipsing binary pulsar (Fruchter,
Stinebring \& Taylor 1988) and the first pulsar with a high-mass
non-degenerate companion (Johnston et al. 1992b). These objects offer
valuable and sometimes profound insight into physics and astrophysics
(e.g. Taylor et al. 1992). Much of the motivation for continued searches
comes from the expectation of finding the unexpected.


\section{The Early Years}
With the announcement of the discovery of the first pulsar, the Molonglo
radio telescope, operated by the University of Sydney, was ideally placed to
follow up on this exciting result. The Cambridge pulsars were discovered with an
array operating at 81.5 MHz, suggesting that pulsars had steep radio
spectra, and the Molonglo telescope operated at the relatively low radio
frequency of 408 MHz. It had large collecting area and so had high
instantaneous sensitivity, necessary to record the rapidly fluctuating
pulsar signals. By November 1968, it had already discovered nine pulsars,
more than half of the then world total of 17 (Large, Vaughan \&
Wielebinski 1968). Included in these was the very important Vela pulsar, the
first to be associated with a supernova remnant.

Initially, astronomers at Parkes concentrated on detailed studies of the
spectrum, polarisation and timing of pulsars, exploiting properties of the
telescope such as frequency versatility, and polarisation and tracking
capability.  These observations were very successful, providing the first
spectra of individual pulses (Robinson et al. 1968), the first observation
of a period glitch (Radhakrishnan \& Manchester 1969) and the genesis of the
now widely accepted `magnetic-pole' model for the emission beam
(Radhakrishnan et al. 1969).

The first succesful search for pulsars at Parkes, reported by Komesaroff et
al. (1973), began in 1973 and discovered eight new pulsars. Because of the
lower instantaneous sensitivity of the Parkes telescope, this survey was
among the first to rely on digital sampling of longer data sets and signal
processing techniques to obtain the necessary sensitivity. Because of its
higher frequency (750 MHz) and the use of multi-channel receivers, this
survey was sensitive to high-DM pulsars. In particular, it discovered the
highly luminous pulsar PSR B1641$-$45, which has a DM of about 480 cm$^{-3}$
pc, the highest known at the time.

\begin{figure}[ht]
\begin{center}
\centerline{\psfig{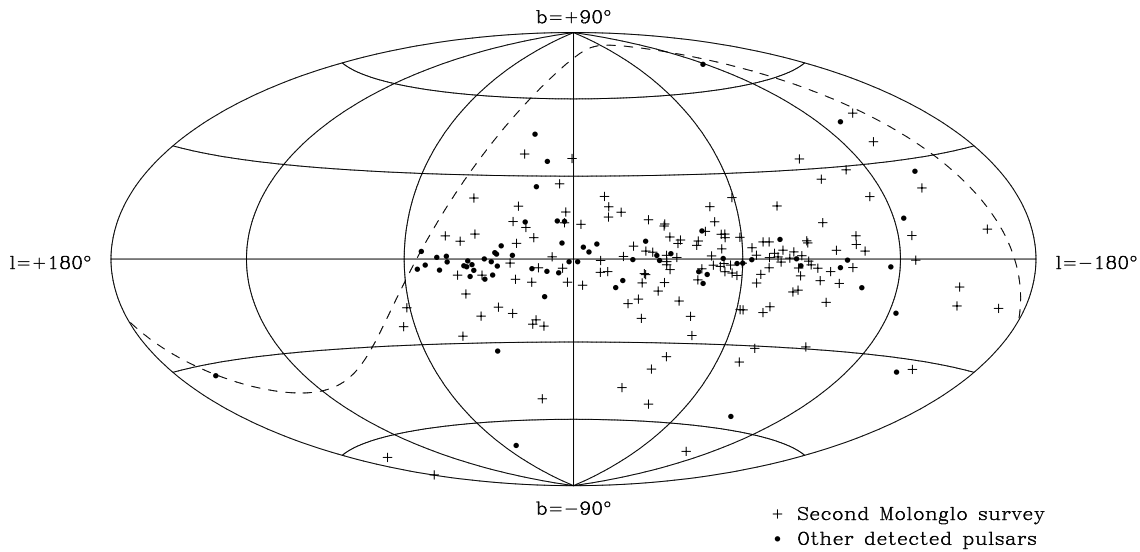}}
\caption{Pulsars detected in the Second Molonglo survey (Manchester et
al. 1978). The dashed line marks declination $+20^{\circ}$, the northern
limit of the survey. }
\label{fg:mol2}            
\end{center}
\end{figure}

In a good example of synergy, the strengths of the Molonglo telescope and
the Parkes telescope were combined to undertake the highly successful Second
Molonglo pulsar survey (Manchester et al. 1978). This survey discovered 154
previously unknown pulsars, more than doubling the number of pulsars known
at the time. The Molonglo telescope was used in a multibeam mode, giving
eight adjacent beams in right ascension which increased the effective
integration time to 44/cos$\,\delta$ sec. Improved front-end amplifiers and
multi-channel receivers were also constructed specifically for this
survey. Candidates from analysis of the Molonglo data were confirmed at
Parkes. The Parkes telescope tracked up the $4^{\circ}$ declination width
of the Molonglo beam with an effective 300-sec integration at
each point. Data were searched in real time about the
candidate parameters, thereby giving an improved declination, period and DM for
confirmed pulsars.  As shown in Fig.~\ref{fg:mol2}, the survey covered the
whole sky south of declination $+20^{\circ}$ and detected a total of 224
pulsars, giving an excellent sample for statistical studies.

The long integration times and wide bandwidths available at Parkes make
possible very sensitive surveys. One such survey was that of McConnell et
al. (1991) which detected the first known extragalactic pulsars, in the
Magellanic Clouds. One of these, PSR J0045$-$7319, until recently the only
known pulsar in the Small Magellanic Cloud, was later shown by Kaspi et
al. (1994) to be in a orbit about an optically identified B-star companion. 

The next major survey to be undertaken at Parkes was the 20-cm survey of
Johnston et al. (1992a). This survey covered a strip along the 
Galactic plane with $270^{\circ} < l < 20^{\circ}$ and $|b| < 4^{\circ}$,
complementing a similar survey of the northern Galactic plane (Clifton et
al. 1992). A bandwidth of 320 MHz centred at 1520 MHz was observed with an
effective integration time per point of 2.5 min, giving a limiting
sensitivity of about 1 mJy for pulsars with period greater than about 50
ms. A total of 100 pulsars were detected by the survey, with 46 being new
discoveries. Included in them was the very interesting eclipsing binary
pulsar PSR B1259$-$63 (Johnston et al. 1992b). This pulsar is in a 3.5-year
highly eccentric orbit around a 10-M$_{\odot}$ Be star SS 2883, and was the
first pulsar known to have a massive non-degenerate companion. Near
periastron, the pulsar passes through the circumstellar disk of the Be star
and is eclipsed for about 30 days. Significant changes in DM and rotation
measure are observed before and after the eclipse, giving information on the
properties of the circumstellar disk (Johnston et al. 1996). 

Although the Johnston et al. (1992a) survey had sensitivity to MSPs at about
the 2.5 mJy level, none was detected. The main reasons for this were the
high dispersion, scattering and background temperature along the Galactic
plane, coupled with the low luminosity of most MSPs. Also, because of their
great age, most disk millisecond pulsars are at large Galactic
$z$-distances, comparable to or larger than the reach of most
surveys. Consequently they have a nearly isotropic distribution on the
sky. These considerations suggested that a lower-frequency search covering a
large area of the sky would be more likely to detect a significant number of
MSPs. The Parkes 70cm survey (Manchester et al. 1996, Lyne et al. 1998) was
designed with these ideas in mind.

\begin{figure}[t]
\begin{center}
\centerline{\psfig{file=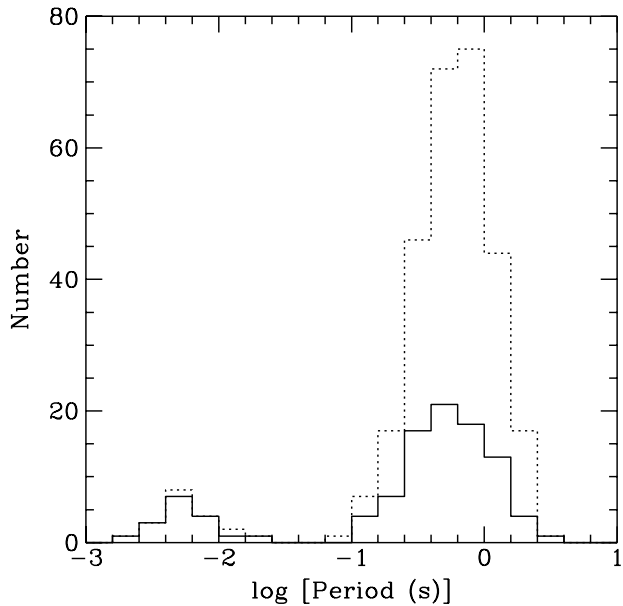,width=120mm}}
\caption{Period distribution of pulsars discovered (full line) and detected
(dotted line) by the Parkes Southern pulsar survey (Lyne et al. 1998)}
\label{fg:sur70_prd}            
\end{center}
\end{figure}

The survey covered the whole sky south of the equator at a frequency of 436
MHz, with a sampling interval of 300 $\mu$s and an observation time per
point of 157 sec, giving it a limiting sensitivity of about 3 mJy. It
detected 298 pulsars of which 101, including 17 MSPs, were previously
unknown. Fig.~\ref{fg:sur70_prd} shows the period distribution of these
pulsars. This figure highlights the fact that MSPs are a different
population, quite distinct from the normal pulsars. As expected, the sky
distribution of MSPs was close to isotropic, whereas the normal pulsars were
clustered along the Galactic Plane. The large number of pulsars detected and
the well defined survey parameters make this an excellent data base for
studies of the Galactic distribution and birthrate of both normal pulsars
and MSPs. Lyne et al. (1998) estimate that there are about 30,000
potentially observable MSPs with 400 MHz luminosity above 1 mJy kpc$^2$ and
a similar number of potentially observable normal pulsars above the same
luminosity limit in the Galaxy. After taking beaming into account, the
corresponding birth rate for normal pulsars is one per 60 to 330 years,
and for MSPs, one per 300,000 years.

\begin{figure}[t]
\begin{center}
\centerline{\psfig{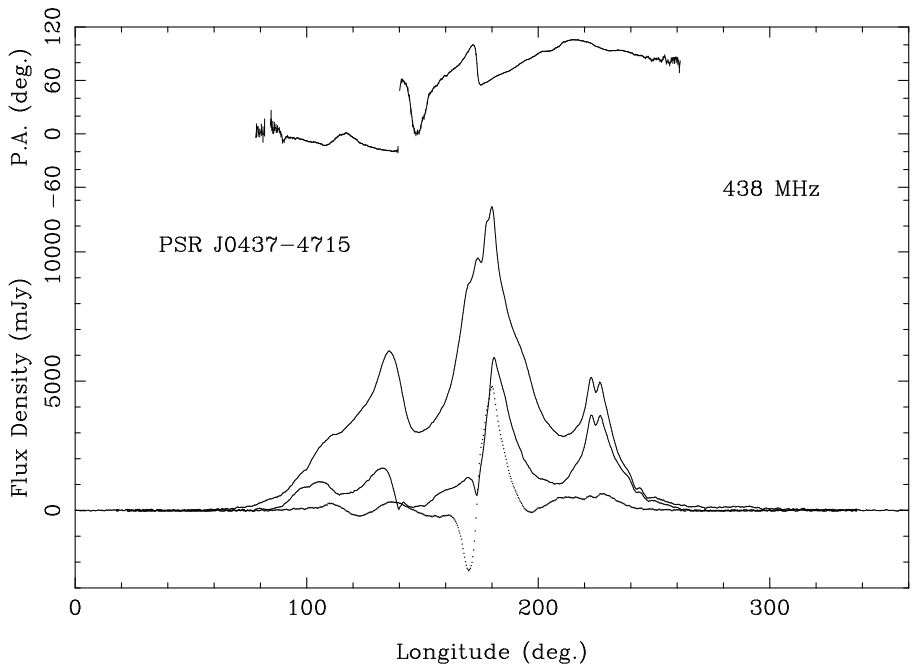}}
\caption{Mean pulse profile and polarisation parameters for PSR J0437$-$4715
(Navarro et al. 1997). The entire pulse period is shown. In the lower part
of the figure, the upper line is the total intensity, Stokes $I$, the other
solid line is the linearly polarised intensity, $L=(Q^2 + U^2)^{1/2}$, and
the dotted line is the circularly polarised intensity, Stokes $V$. The line
in the upper part is the position angle of the linearly polarised
component. }
\label{fg:0437}            
\end{center}
\end{figure}

Probably the most interesting pulsar discovered by the Parkes Southern
survey was PSR J0437$-$4715, by far the nearest and strongest millisecond
pulsar known (Johnston et al. 1993). This pulsar has a period of 5.75 ms, is
a 5.74-day binary orbit with a companion of mass $\sim 0.3$ M$_{\odot}$, and
has a mean flux density at 430 MHz of more than 500 mJy. The strength of
this pulsar makes possible very precise measurements of its polarisation and
timing properties. As shown in Fig.~\ref{fg:0437}, the pulsar has a very wide and
complex profile covering more than 80\% of the period with at least 12
identifiable pulse components and high polarisation, both linear and
circular (Navarro et al. 1997). The position angle variation is complex,
suggesting that the usual assumption of dipole magnetic field lines in the
pulsar magnetosphere is not valid. Timing observations (Sandhu et al. 1997)
have given the pulsar position with a precision of 50 $\mu$as, the proper
motion at the 2000-$\sigma$ level, and a value for the annual parallax, $5.6
\pm 0.8$ mas. This parallax corresponds to a distance for the pulsar about
30\% larger than the value derived from the DM. The large proper motion
($\sim 140$ mas yr$^{-1}$) results in an apparent acceleration of the pulsar,
accounting for about 80\% of the observed period derivative. It also changes
the inclination of the pulsar orbit to the line of sight, resulting in a
secular change in the projected size of the orbit. This change has been
detected at the 20-$\sigma$ level, enabling a limit to be placed on the
inclination angle of the orbit, $i < 43^{\circ}$, and hence improving the
determination of the mass of the companion.

With hindsight, another very interesting pulsar discovered in this survey
was PSR J2144$-$3933, originally believed to have a pulse period of
2.84~s. As part of a study of the pulse-to-pulse fluctuation properties of
pulsars, Matthew Young realised that the true period of this pulsar was
8.51~s, by far the longest period known (Young, Manchester \& Johnston
1999). This long period places the pulsar beyond the `death line' of most
models for the pulse emission mechanism. The pulsar also has a very narrow
pulse, less than $1^{\circ}$ of longitude. If this is typical of such
long-period pulsars, they could form a large fraction of the total Galactic
population.


\section{The Parkes Multibeam Pulsar Survey}
The Parkes multibeam receiver, while primarily designed for HI surveys
(Staveley-Smith et al. 1996), is a superb instrument for pulsar surveys. Its
13 beams allow the sky to be covered roughly 13 times as fast, or
alternatively, much longer to be spent on a given point. Also, its receivers
have excellent sensitivity, with an average system noise of only 21 K. With
major contributions from Jodrell Bank Observatory and Osservatorio
Astronomica di Bologna, a filterbank system capable of handling the data
from the 13 beams was installed at Parkes in early 1997 and the Parkes
multibeam pulsar survey commenced in August 1997. The survey is covering the
region $260^{\circ} < l < 50^{\circ}$ with $|b| < 5^{\circ}$. The filterbank
has 96 3-MHz channels for each polarisation of each beam and all outputs are
one-bit digitised at 250 $\mu$s intervals and recorded to tape. Each
pointing of the 13 beams is of 35 min duration, giving a sensitivity for
long-period pulsars away from the hot regions of the Galactic background of
about 0.15 mJy. This is about seven times better than the previous best
survey of this type (Johnston et al. 1992a), and so a large increase in the
number of detected pulsars was expected.

\begin{figure}[ht]
\begin{center}
\centerline{\psfig{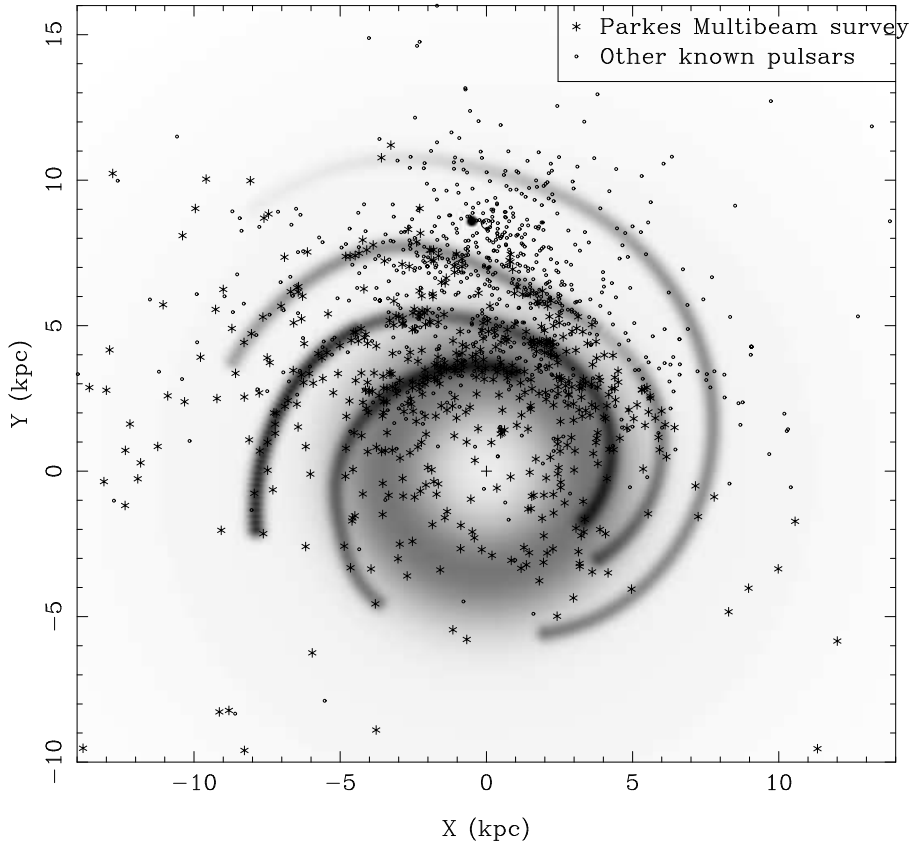}}
\caption{Distribution of known pulsars projected on to the Galactic
plane. The position of the Sun is marked by $\odot$ and the Galactic
Centre by +. The distribution of interstellar free electrons according to the
Taylor \& Cordes (1993) model is shown as a grey-scale. The dark spot near
the Sun is the Gum Nebula. }
\label{fg:pm_xy}            
\end{center}
\end{figure}

These expectations have already been fully realised. With about 80\% of the
survey completed, more than 570 previously unknown pulsars have been
discovered, making this by far the most successful pulsar survey ever. When
finished, the survey will come close to doubling the number of known
pulsars. Fig.~\ref{fg:pm_xy} shows the distribution of known pulsars
projected on to the Galactic plane, where distances have been computed using
the Taylor \& Cordes (1993) electron density model. In contrast to the
previously known pulsars which are clustered around the Sun, many of the
multibeam pulsars are at large distances, with some apparently on the other
side of the Galactic Centre. There is some indication of a deficit of
detected pulsars within a couple of kpc of the Galactic Centre. The electron
density model is not well determined at these large distances though, and
the distances may have systematic biases. Also, many of the multibeam
pulsars are concentrated in spiral arms, but this may simply be a result of
the increased model electron density in the arms. The multibeam sample will
be important in helping to refine the electron density model.
 
\begin{figure}[ht]
\begin{center}
\centerline{\psfig{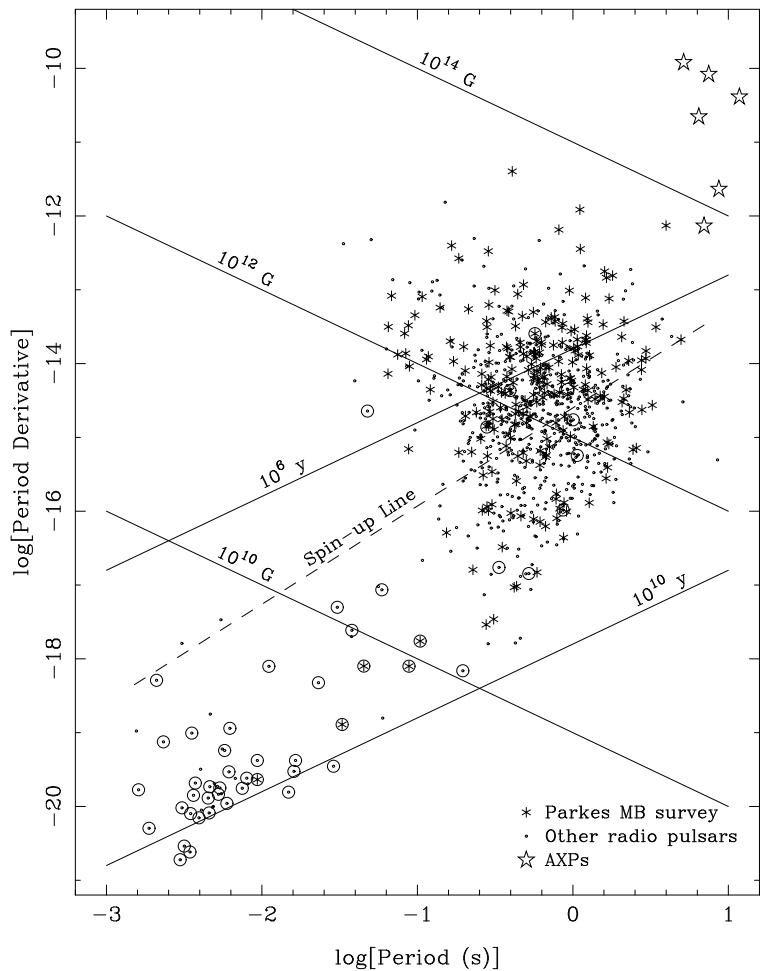}}
\caption{Distribution of pulsars and anomalous X-ray pulsars (AXPs) in the
$P - \dot P$ plane. Binary systems are indicated by a circle around the
point. Lines of constant pulsar characteristic age, $\tau_c = P/(2\dot P)$ and
surface dipole magnetic field strength, $B_s \propto (P\dot P)^{1/2}$ are
indicated. The spin-up line, representing the minimum period attainable by
accretion from a binary companion, is also shown.}
\label{fg:pm_ppdot}            
\end{center}
\end{figure}

As shown by Fig.~\ref{fg:pm_ppdot}, pulsars detected by the multibeam survey
are on average much younger than previously known pulsars.  They include the
three pulsars with the strongest known surface dipole magnetic fields
(Camilo et al. 2000a). One of these pulsars, PSR J1119$-$6127, has a
characteristic age of only 1700 years and is associated with what appears to
be a previously uncatalogued supernova remnant (Crawford et
al. 2000). Another, PSR J1814$-$1744, has the relatively long period of 3.97
sec, but a very rapid spin-down rate giving it an implied surface field
strength of $5.5 \times 10^{13}$ G. These parameters place the pulsar near
the so-called `anomalous X-ray pulsars' (AXPs) on the $P - \dot P$ plane
(Fig.~\ref{fg:pm_ppdot}). AXPs are believed to be slowly rotating neutron
stars, but they have no detectable radio emission. On the other hand, PSR
J1814$-$1744 has no detectable X-ray emission (Pivovaroff, Kaspi \& Camilo
2000). The reason(s) for these very different properties are not well
understood.

Only one millisecond pulsar has been discovered so far. Although this is
somewhat surprising given the parameters of the survey, there are several
possible contributing factors. The search so far has been concentrated at
low Galactic latitudes where dispersion and scattering are large. This
limits the maximum distance at which typical MSPs can be detected and so the
volume of the Galaxy searched so far is relatively small. Another
contributing factor is that only `unaccelerated' searches have been
performed so far. Especially with the rather long observation time of this
survey, this limits our senstivity to millisecond pulsars, most of which are
in binary systems. A third factor is that algorithms for dealing with
interference were not optimal in the early stages of processing. This mainly
affects MSP detections, since most of the searched frequency space
corresponds to millisecond periods. All of these factors are being overcome
and we expect to detect more MSPs with future observations and data
processing.

Eight of the newly discovered pulsars are members of binary systems. Five of
these are in near-circular orbits with companions which are probably white
dwarfs (Camilo et al. 2000c). These systems differ from most known white
dwarf binaries. Except for the one MSP detected, the pulsar periods are
relatively long, lying between 45 and 90 ms, and the companions are heavy
with minimum mass between 0.15 and 0.9 M$_{\odot}$.

One of the binary pulsars, PSR J1811$-$1736, is in a highly eccentric
18-day orbit and is very probably a double neutron-star system, the first to
be discovered in the southern hemisphere (Lyne et al. 2000). In contrast to
PSR J1811$-$1736 which has a characteristic age $\sim 10^9$ yr, PSR
J1141-6545 is a young pulsar ($\tau_c \sim 1.5 \times 10^6$ yr) in a much
tighter ($P_b \sim 4.1$ h) and eccentric orbit (Kaspi et
al. 2000). Precession of the longitude of periastron has been observed for
this system, and interpreting this as due to the effects of general
relativity gives a value for the total mass of the system of $2.300 \pm 0.012$
M$_{\odot}$. The pulsar and orbit properties suggest that the companion is a
heavy white dwarf formed {\it before} the supernova explosion that created
the pulsar. This is unusual. In most binary systems the neutron star is
formed from the heavier binary companion which evolves faster.

\begin{figure}[ht]
\begin{center}
\centerline{\psfig{file=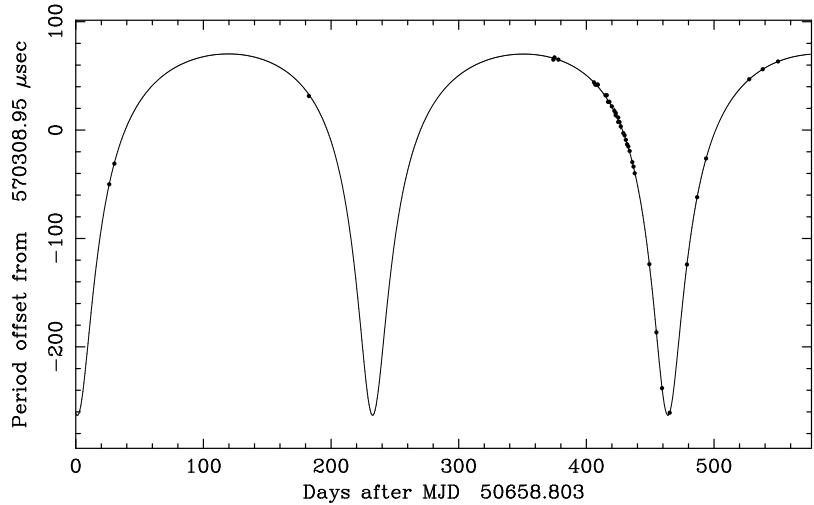,width=120mm}}
\caption{Variations in apparent solar-system barycentric period of PSR
J1740$-$3052. The fitted line is for a binary model with orbital period
230.0 days and eccentricity 0.579 (Stairs et al. 2000).}
\label{fg:1740_per}            
\end{center}
\end{figure}

As shown in Fig.~\ref{fg:1740_per}, PSR J1740$-$3052 is in a highly
eccentric long-period orbit. The interesting thing about this system is that
the minimum companion mass is 11 M$_{\odot}$, implying that the companion is
either a massive star or a black hole. Unfortunately the pulsar lies close
to the direction of the Galactic Centre and probably at about the same
distance, so optical searches for the companion are unlikely to be
productive. However, 2.2 $\mu$m infrared observations with the Siding Spring
2.3-m telescope and the Anglo-Australian Telescope have revealed a
K-supergiant star whose position agrees with that of the pulsar to better
than 0.3 arcsec (Stairs et al. 2000). The infrared spectrum of this star
shows Brackett-$\gamma$ emission, consistent with the presence of a compact
binary companion, and the star's colours are consistent with a distance
comparable to that of the pulsar. Furthermore, DM and rotation measure
changes were observed over the last periastron passage, in February
2000. All of these observations point toward this star being the binary
companion. However, there a couple of puzzling features. The pulsar comes to
within 1.25 stellar radii of the companion star at periastron. One might
expect the radio emission to be eclipsed by the stellar atmosphere or wind,
but no eclipses are observed. Also, it should raise large tides on the
companion, causing a large precession in the longitude of periastron. This
is not observed. Either we do not understand winds and tides in supergiant
stars very well, or all the other observations are misleading and the
companion is really a black hole. Although the latter is an attractive
option (this would be the first known neutron star -- black hole system),
the former is more likely.

\section{Millisecond Pulsars in 47 Tucanae}
In terms of finding millisecond pulsars, the globular cluster 47 Tucanae has
proven to be a gold-mine. The cluster is massive ($\sim 10^6$ M$_{\odot}$),
relatively nearby with a distance from the Sun of about 5 kpc, and has a
very dense core with a central density of $\sim 10^5$ M$_{\odot}$
pc$^{-3}$. These properties are favourable for the formation in the core of
neutron-star binary systems by exchange interactions (e.g. Rasio, Pfahl \&
Rappaport 2000) and the detection of millisecond pulsars formed by
subsequent accretion spin-up of the captured neutron stars. Parkes searches
at 430 and 640 MHz MHz in the early 1990's (Manchester et al. 1990, 1991,
and Robinson et al. 1995) resulted in the detection of 11 millisecond
pulsars in 47 Tucanae, already a record number of pulsars for one
cluster. These pulsars all had very short periods, in the range 2.1 to 5.8
ms, and three of them were known to be members of binary systems. One, 47
Tuc J, had the very short orbital period of 2.9 h and showed evidence for
eclipses of the pulsar signal. Most of these pulsars are very weak and
several were only detected occasionally, when interstellar scintillation
raised their flux density to a detectable level. Because of this only two of
the 11 pulsars, C and D, both isolated (non-binary) pulsars, had coherent
timing solutions and hence accurate positions, periods and slow-down rates.

\begin{figure}[ht]
\begin{center}
\centerline{\psfig{file=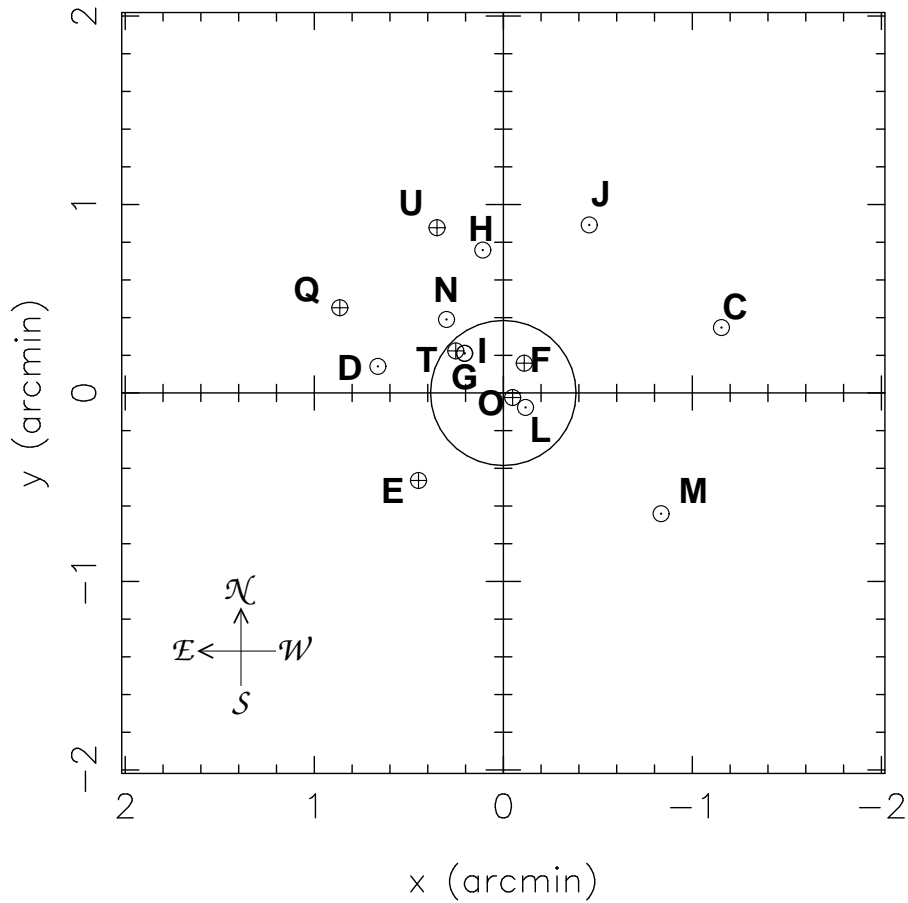,width=120mm}}
\caption{Positions relative to the cluster centre of the 15 pulsars in 47
Tucanae with coherent timing solutions. The large circle indicates the size
of the cluster core. (Freire et al. 2000).}
\label{fg:47tuc_posn}            
\end{center}
\end{figure}

Matters remained like this for several years, until Camilo et al. (2000b)
exploited the high sensitivity of the multibeam system and sophisticated
data processing techniques to discover a further nine MSPs in the
cluster, all members of binary systems! The complete dominance of binary
pulsars in these latest discoveries shows that the low proportion
binary systems in the earlier results was purely a selection effect, and
that most pulsars in globular clusters (at least 47 Tucanae, but probably
all clusters) are binary. Again, all of these pulsars had very
short periods, the longest being 7.6 ms. 47 Tuc R has the shortest orbital
period known for any radio pulsar, 96 min, and also shows evidence for
eclipses. All of the known binary systems in 47 Tucanae have very low-mass
companions, $\lapp 0.2$ M$_{\odot}$.

\begin{figure}[ht]
\begin{center}
\centerline{\psfig{file=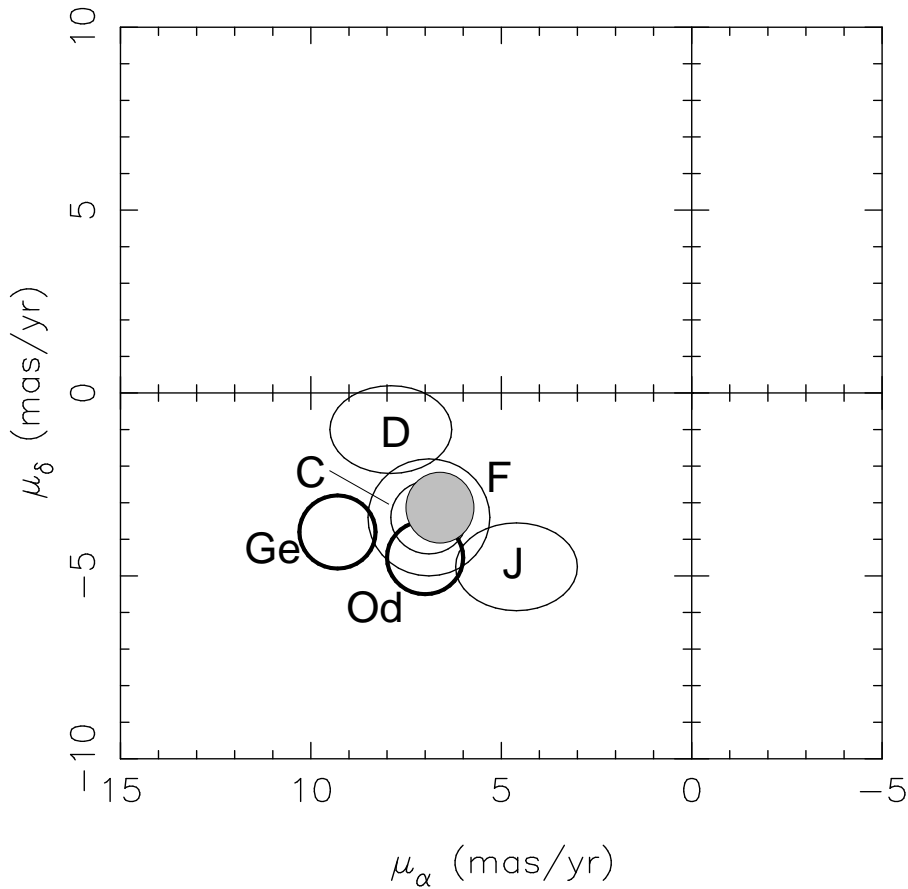,width=120mm}}
\caption{Proper motions of four pulsars in 47 Tucanae (Freire et
al. 2000). The filled ellipse represents a weighted average of the pulsar
results and its uncertainty. Also shown are two determinations
of the cluster proper motion from Hipparcos data by Geffert et al. (1997)
and Odenkirchen et al. (1997).}
\label{fg:47tuc_pm}            
\end{center}
\end{figure}

Not only has the multibeam system allowed detection of more pulsars in 47
Tucanae, it has also given us a much higher detection rate on the previously
known pulsars. This has allowed timing solutions to be obtained for a
further 13 pulsars giving us a much improved knowledge of the pulsar
parameters (Freire et al. 2000). This in turn makes possible a number of
interesting studies of cluster properties. Fig.~\ref{fg:47tuc_posn} shows
that the pulsars are concentrated in the central region of the cluster; the
core radius is 23 arcsec but the tidal radius is much larger, about 40
arcmin. This indicates that they are in dynamical equilibrium with other
cluster stars, with their larger than average mass giving them a smaller
than average velocity in the cluster and hence confining them to the central
region. Interestingly, a cumulative histogram of number of pulsars versus
perpendicular radius from the cluster centre is linear within the
uncertainties. This implies that stars with mass similar to that of a neutron
star are the dominant stellar species in the core region. Since the maximum
mass of main-sequence stars is about 0.8 M$_{\odot}$, these may be
unseen neutron stars or binary systems consisting of two heavy main sequence
stars. The existence of such binary systems is suggested by the observation
of `blue straggler' stars in the cluster core (Gilliand et al. 1998); these
stars are believed to be formed by the coalescence of such binary systems. 

The positions obtained from timing observations have very small
uncertainties, typically $\lapp 1$ mas. Given the improved parameters for
cluster pulsars obtained from the recent observations, Freire et al. (2000)
were able to go back and reanalyse the data from the early 1990's to obtain
two more coherent timing solutions, giving four in all. Comparison of
positions from these solutions with those from analyses of recent data
allowed determination of the proper motion of the four pulsars. These
results are shown in Fig.~\ref{fg:47tuc_pm}. The observed proper motions are
dominated by the motion of the cluster as a whole, not by motions of pulsars
relative to the cluster. Their weighted mean already gives a more precise
value for the cluster proper motion than that obtained from Hipparchos data
and future observations will further improve the reliability of this value.

\begin{figure}[ht]
\begin{center}
\centerline{\psfig{file=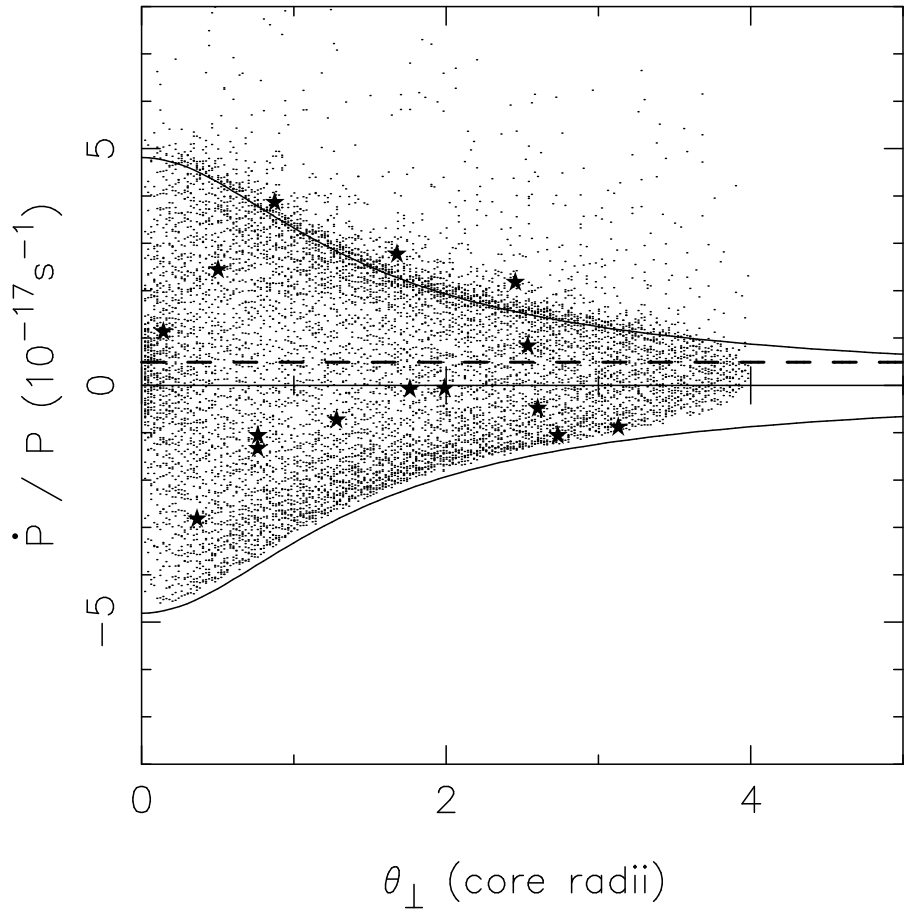,width=120mm}}
\caption{Observed $\dot P/P$ ratios for pulsars in 47 Tucanae as a function
of their radial distance from the cluster core (Freire et al. 2000). The
dashed line is the mean observed value of $\dot P/P$. The curved solid lines
give the maximum acceleration based on a King model for the cluster density
distribution and the dots represent accelerations of a random sample of
pulsars having the mean acceleration and distributed through the cluster
with an $r^{-2}$ density distribution. }
\label{fg:47tuc_accel}            
\end{center}
\end{figure}

As mentioned in the Introduction, the long-term or secular intrinsic period
derivative of all pulsars is believed to be positive. However, observed
period derivatives for MSPs near the core of globular clusters are sometimes
negative. This is attributed to an acceleration of the pulsar in the cluster
gravitational potential dominating over any intrinsic period
derivative. Pulsars on the far side of the cluster will be accelerating
toward us and hence will appear to be speeding up. Fig.~\ref{fg:47tuc_accel}
shows the observed $\dot P/P$ ratios along with a model prediction of the
probability of a given acceleration being observed. The observed points are
consistent with these predictions provided the central velocity dispersion
in the cluster is more than 12 \kms (Freire et al. 2000).

These results are already impressive. However, with continuing observations
using (already installed) filterbanks giving higher time resolution, new and
improved results can be expected from the Parkes observations of 47
Tucanae. Already, two more pulsars have been discovered bringing the total
known in the cluster to 22. Existing timing solutions will improve and new
timing solutions will be obtained, giving improved proper motions
and cluster accelerations. There is no doubt that further study of 47
Tucanae will be rewarding.


\section{Conclusions}
In this review, I have highlighted the major pulsar searches 
undertaken using the Parkes 64-m radio telescope. Numerous other
smaller-scale searches have taken place over the years, many of which have
produced important results. A recent notable example is the survey being
undertaken by Russell Edwards and his colleagues from Swinburne University
of Technology using the multibeam receiver. The Swinburne survey complements
the Parkes multibeam survey, covering higher latitudes with
$5^{\circ}<|b|<15^{\circ}$ and has discovered 55 pulsars including eight
MSPs (Edwards 2000).

All in all, the Parkes telescope has discovered just under two thirds of all
known pulsars, nearly twice as many as all the other telescopes in the world
combined! And the most successful of these surveys, the Parkes multibeam
survey is not yet finished. By a large margin, the Parkes telescope is the
most successful telescope in the world at finding pulsars. This success is
due to several factors, principally the large size of the telescope, its
southern location (making it ideal for Galactic surveys) and the excellent
performance and innovative design of the receiver systems installed on it. I
have to say that the stamina of the astronomers responsible for these
surveys is also a significant factor.




\section*{Acknowledgements}
There are two very important acknowledgements to be made. The first is to
the people who designed, built, maintained and developed the Parkes
telescope and its receivers over its 40-year history. The second is to my
colleagues who have played a major role in most of the work described
here. Without the skill and dedication of these people, this review could
never have been written. I also thank Simon Johnston for a careful reading
of the manuscript and several helpful suggestions. The Parkes radio
telescope is part of the Australia Telescope which is funded by the
Commonwealth of Australia for operation as a National Facility managed by
CSIRO.


\section*{References}



\reference
Alpar, M.~A., Cheng, K.~S., \& Pines, D. 1989, ApJ, 346, 823

\reference
Anderson, S.~B. 1992,
PhD thesis, California Institute of Technology

\reference
Backer, D.~C., Kulkarni, S.~R., Heiles, C., Davis, M.~M., \& Goss, W.~M. 1982,
  Nature, 300, 615

\reference
Bhattacharya, D. \& {van den Heuvel}, E. P.~J. 1991, Phys. Rep., 203, 1

\reference
Camilo, F.~M., Kaspi, V.~M., Lyne, A.~G., Manchester, R.~N., Bell, J.~F.,
  D'Amico, N., McKay, N. P.~F., \& Crawford, F. 2000a, ApJ,
in press

\reference
{Camilo}, F., {Lorimer}, D.~R., {Freire}, P., {Lyne}, A.~G., \& {Manchester},
  R.~N. 2000b, ApJ, 535, 975

\reference
Camilo, F. {et al.}  2000c, ApJ, submitted

\reference
Clifton, T.~R., Lyne, A.~G., Jones, A.~W., McKenna, J., \& Ashworth, M. 1992,
  MNRAS, 254, 177

\reference
Crawford, F., Gaensler, B., Kaspi, V.~M., Manchester, R.~N., Camilo, F., Lyne,
  A.~G., \& Pivovaroff, M.~J. 2000, ApJ, submitted

\reference
Edwards, R.~T. 2000, in Pulsar Astronomy - 2000 and Beyond, {IAU} Colloquium 177, 
ed.\ M. Kramer, N. Wex, \& R. Wielebinski, Astronomical Society of the Pacific, 33

\reference
Frail, D.~A., Weisberg, J.~M., Cordes, J.~M., \& Mathers, C. 1994, ApJ, 436, 144

\reference
Freire, P.~C., Camilo, F., Lorimer, D.~R., Lyne, A.~G., \& Manchester, R.~N.
  2000, in Pulsar Astronomy - 2000 and Beyond, {IAU} Colloquium 177, 
ed.\ M. Kramer, N. Wex, \& R. Wielebinski, Astronomical Society of the Pacific, 87

\reference
Fruchter, A.~S., Stinebring, D.~R., \& Taylor, J.~H. 1988, Nature, 333, 237

\reference
{Geffert}, M., {Hiesgen}, M., {Colin}, J., {Dauphole}, B., \& {Ducourant}, C.
  1997, ESA Symp. 402: Hipparcos - Venice '97, 402, 579

\reference
{Gilliland}, R.~L., {Bono}, G., {Edmonds}, P.~D., {Caputo}, F., {Cassisi}, S.,
  {Petro}, L.~D., {Saha}, A., \& {Shara}, M.~M. 1998, ApJ, 507, 818

\reference
Hamilton, T.~T., Helfand, D.~J., \& Becker, R.~H. 1985, AJ, 90, 606

\reference
{Han}, J.~L., {Manchester}, R.~N., \& {Qiao}, G.~J. 1999, MNRAS, 306, 371

\reference
Hewish, A., Bell, S.~J., Pilkington, J. D.~H., Scott, P.~F., \& Collins, R.~A.
  1968, Nature, 217, 709

\reference
Hulse, R.~A. \& Taylor, J.~H. 1974, ApJ, 191, L59

\reference
Johnston, S. {et al.}  1993, Nature, 361, 613

\reference
Johnston, S., Lyne, A.~G., Manchester, R.~N., Kniffen, D.~A., D'Amico, N., Lim,
  J., \& Ashworth, M. 1992a, MNRAS, 255, 401

\reference
Johnston, S., Manchester, R.~N., Lyne, A.~G., Bailes, M., Kaspi, V.~M., Qiao,
  G., \& D'Amico, N. 1992b, ApJ, 387, L37

\reference
Johnston, S., Manchester, R.~N., Lyne, A.~G., D'Amico, N., Bailes, M.,
  Gaensler, B.~M., \& Nicastro, L. 1996, MNRAS, 279, 1026

\reference
Kaspi, V.~M., Johnston, S., Bell, J.~F., Manchester, R.~N., Bailes, M.,
  Bessell, M., Lyne, A.~G., \& D'Amico, N. 1994, ApJ, 423, L43

\reference
Kaspi, V.~M. {et al.}  2000, ApJ, in press

\reference
Komesaroff, M.~M., Ables, J.~G., Cooke, D.~J., Hamilton, P.~A., \& McCulloch,
  P.~M. 1973, Astrophys. Lett., 15, 169

\reference
Large, M.~I., Vaughan, A.~E., \& Mills, B.~Y. 1968, Nature, 220, 340

\reference
Large, M.~I., Vaughan, A.~E., \& Wielebinski, R. 1968, Nature, 220, 753

\reference
Lyne, A.~G., Brinklow, A., Middleditch, J., Kulkarni, S.~R., Backer, D.~C., \&
  Clifton, T.~R. 1987, Nature, 328, 399

\reference
Lyne, A.~G. {et al.}  2000, MNRAS, 312, 698

\reference
Lyne, A.~G. {et al.}  1998, MNRAS, 295, 743

\reference
McConnell, D., McCulloch, P.~M., Hamilton, P.~A., Ables, J.~G., Hall, P.~J.,
  Jacka, C.~E., \& Hunt, A.~J. 1991, MNRAS, 249, 654

\reference
Manchester, R.~N. {et al.}  1996, MNRAS, 279, 1235

\reference
Manchester, R.~N., Lyne, A.~G., D'Amico, N., Johnston, S., Lim, J., \& Kniffen,
  D.~A. 1990, Nature, 345, 598

\reference
Manchester, R.~N., Lyne, A.~G., Robinson, C., D'Amico, N., Bailes, M., \& Lim,
  J. 1991, Nature, 352, 219

\reference
Manchester, R.~N., Lyne, A.~G., Taylor, J.~H., Durdin, J.~M., Large, M.~I., \&
  Little, A.~G. 1978, MNRAS, 185, 409

\reference
Navarro, J., Manchester, R.~N., Sandhu, J.~S., Kulkarni, S.~R., \& Bailes, M.
  1997, ApJ, 486, 1019

\reference
{Odenkirchen}, M., {Brosche}, P., {Geffert}, M., \& {Tucholke}, H.~J. 1997, New
  Astronomy, 2, 477

\reference
Pivovaroff, M., Kaspi, V.~M., \& Camilo, F. 2000, ApJ, in press

\reference
Radhakrishnan, V., Cooke, D.~J., Komesaroff, M.~M., \& Morris, D. 1969, Nature,
  221, 443

\reference
Radhakrishnan, V. \& Manchester, R.~N. 1969, Nature, 222, 228

\reference
{Rasio}, F.~A., {Pfahl}, E.~D., \& {Rappaport}, S. 2000, ApJ, 532, 47

\reference
Rickett, B.~J. 1990, Ann. Rev. Astr. Ap., 28, 561

\reference
Robinson, B.~J., Cooper, B. F.~C., Gardiner, F.~F., Wielebinski, R., \&
  Landecker, T.~L. 1968, Nature, 218, 1143

\reference
Robinson, C.~R., Lyne, A.~G., Manchester, A.~G., Bailes, M., D'Amico, N., \&
  Johnston, S. 1995, MNRAS, 274, 547

\reference
Sandhu, J.~S., Bailes, M., Manchester, R.~N., Navarro, J., Kulkarni, S.~R., \&
  Anderson, S.~B. 1997, ApJ, 478, L95

\reference
Staelin, D.~H. \& Reifenstein, E.~C. 1968, Science, 162, 1481

\reference
Stairs, I.~H. {et al.}  2000, MNRAS, submitted

\reference
Staveley-Smith, L. {et al.}  1996, PASA, 13, 243

\reference
Taylor, J.~H. \& Cordes, J.~M. 1993, ApJ, 411, 674

\reference
Taylor, J.~H., Wolszczan, A., Damour, T., \& Weisberg, J.~M. 1992, Nature, 355,
  132

\reference
Wang, N., Manchester, R.~N., Pace, R., Bailes, M., Kaspi, V.~M., Stappers,
  B.~W., \& Lyne, A.~G. 2000, MNRAS, in press

\reference
Wolszczan, A., Kulkarni, S.~R., Middleditch, J., Backer, D.~C., Fruchter,
  A.~S., \& Dewey, R.~J. 1989, Nature, 337, 531

\reference
{Young}, M.~D., {Manchester}, R.~N., \& {Johnston}, S. 1999, Nature, 400, 848






\end{document}